\documentclass[prl,aps,twocolumn,showpacs,nopreprintnumbers]{revtex4}
\usepackage{bm,bbm,amsmath,amssymb,subeqnarray,graphicx}
\usepackage{epsf}
\usepackage{graphicx,epsfig}
\usepackage{bm}
\usepackage{latexsym,float}
\usepackage{subfig}
\usepackage{dcolumn}
\usepackage{ulem}
\usepackage{gensymb}

\usepackage{amssymb}
\usepackage{color}
\usepackage{amsmath}
\usepackage{graphicx}
\usepackage{bm}
\usepackage{epsfig}

\begin{document}

\title{ Collimation and acceleration of monoenergetic electron beams in Laser Plasma accelerators }
\author{Ravindra Kumar}
\affiliation{Department of Physics, I.I.T. Kanpur, Kanpur
208016(INDIA)}
\author{V.Ravishankar}
\affiliation{Department of Physics, I.I.T. Kanpur, Kanpur
208016(INDIA)}

\begin{abstract}
Motivated by rapid advances in plasma based accelerators,
 we propose a simple mechanism for the production of highly collimated quasi monochromatic
 electrons beams. By considering  Compton scattering of  electrons (in the plasma) with virtual photons --
 through an effective interaction --- we demonstrate angular collimation with a divergence less than $3~ mrad$, quasi monoenergetic nature of the beam
  and an energy gain of $O (1 GeV/cm)$, in laser induced accelerators.
  
\end{abstract}

\maketitle
 Collective effects in   laser plasma interaction  play a
significant role in table top particle acceleration.   Dawson and Tajima \cite{dt} first showed,
through computer simulations, that a high intensity laser shone on a dense plasma
( $n_0  \sim 10^{18} cm^{-3}$ ) could accelerate the electrons
in the plasma up to a GeV over a short distance of about a centimetre. Subsequently, significant progress has been made
 both experimentally and theoretically \cite{modena,umstadter,moore,science,faure,mangles,geddes,malka,muggli}
  addressing, mainly, two issues: (i) acceleration to high energies, and (ii) achieving
a good collimation so that they can be of utility in  high energy experiments and other applications \cite{muggli}.
While beam accelerations up to a few GeV had been achieved prior to 2004, the spatial divergence of the beam ranged from a few
to tens of degrees with a Maxwellian
energy spectrum  \cite{modena,umstadter,moore}  and a large energy spread $(\Delta E/E) \sim 100\%$. Monoenergetic
beams with small spatial divergence  were achieved first by three groups\cite{faure,geddes,mangles},
henceforth referred to as I, II and III respectively. In I and II  collimation was achieved by carefully choosing
 laser parameters and plasma density   whereas III used a preformed plasma channel to achieve the same.
 Unlike in the earlier experiments, the distribution was non Maxwellian in all the three cases. A few more salient points:
 In I  the electron beam spectrum contained $Q \sim 2 \pm 0.5$ nC of charge with an  energy $E \sim 170 \pm 20$ MeV
 within the collimation angle  $\Delta\theta =10$ mrad.  II reports  $Q\sim 22$ pC of charge
 with $\Delta\theta  \lesssim 50$ mrad. Finally, III found  $Q \sim 0.32 $ nC of
 charge with $E\sim 86 \pm 1.8$ MeV  and $\Delta\theta \sim 3 $ mrad.
Promising that these results are  -- for future accelerator designs and experiments,  theoretical analyses have so far been computation intensive, employing  PIC (three dimensional
particle-in-cell) simulations \cite{dawson} which have been used e.g., in \cite{faure,science},   and Vlasov models \cite{ghizzo}.  Depending on the regime of interest, the acceleration mechanism is attributed to either laser wake field acceleration (in the linear regime),  or stimulated Raman scattering (in the nonlinear regime \cite{joshi2}. In this letter, in contrast to the above mentioned methods, we take an effective field theory approach to unravel the physics behind Plasma acceleration. We study acceleration via  Compton scattering of photons by electrons  in the medium, by explicitly incorporating the collective nature of the plasma {\it at every level}.  We hope to show that the essential physics behind plasma accelerators can be captured by this approach and establish  collimation and the monoenergetic nature of the scattered electrons. The attainment of high energies $O(\sim 1 GeV)$ over  a distance of about a cm,  will be addressed rather qualitatively, and will be treated more completely  in a future work. \\
\noindent {\bf The Effective Interaction}: In the radiation gauge $A^0=0$ which we employ throughout, the constitutive Maxwell equations, which incorporate
electric response of the plasma, have the form

 \begin{eqnarray}
 4\pi\rho (\omega ,\vec k) = -\omega{\varepsilon _l}(\omega ,\vec k)\vec{k}\cdot \vec{A}; \nonumber \\
j_i(\omega,\vec k) =  [\omega ^2\varepsilon _l(\omega,\vec k){\cal P}^l_{ij}
 +({\omega ^2}{\varepsilon_t}(\omega,\vec k)- { {\vec k}^2}) {\cal P}^t_{ij}]{A_j}
\end{eqnarray}
where,  ${\cal P}^{l}_{ij} = k_ik_j/\vec{k}^2;  {\cal P}^t_{ij} = (\delta_{ij} -{\cal P}^l_{ij})$ are the projection operators for  longitudinal and transverse modes respectively and     ${\varepsilon _l}$ and ${\varepsilon _t}$ are the corresponding permittivities. In the collisionless  limit and in the regime
$\frac{\omega}{kv_T}>> 1$ relevant to us, we obtain the well known relations $ {\omega ^2} -
{a_{l,t}}{\left| k \right|^2} = {\omega_p^2}$  where, ${a_l} =
3v_T^2~;~{a_t} = 1 +v_T^2$ in terms of the plasma frequency $\omega_p = \sqrt{4\pi
n_0e^2/m}$  and the thermal velocity of the electron $v_T =
\sqrt{T/m}$. T is the plasma temperature and m is the electronic mass.

 The  lagrangian of the system including the medium
responses can be written  as (we employ the units $\hbar =c = k_B=1 $),
$\mathcal{ L} =
\mathcal{L}_{Dirac}+\mathcal{L}_{Int}+\mathcal{L}_{Field}$, where
$
 \mathcal{ L}_{Dirac} = \bar \psi (x)(i\gamma ^\mu  \partial _\mu   - m)\psi (x); ~
 \mathcal{ L}_{Int} = -e \bar \psi (x)\gamma ^\mu  \psi (x)A_\mu (x)$ retain their usual forms while the field part gets modified to
\begin{equation}
\mathcal{ L}_{Field} = \frac{1}{2}[{\cal P}^l_{ij}E_i {D}^l_j +
 {\cal P}^t_{ij} E_i {D}^t_{j}
-{\vec B^2}]\label{efflag}
\end{equation}
where, $ {D}^{l,t}_i(\vec r,t)=\int{d^3{r^\prime} dt^\prime}{{\varepsilon}^{l,t}(t-t^\prime,\vec r-\vec r^\prime) E_i(t^\prime,r^\prime)}$

The corresponding hamiltonian for the field reads \cite{landau,harris}
\begin{eqnarray}
\mathcal{H}_{Field} &=& \frac{1}{2}[({\delta _{ij}} -
\frac{{{k_i}{k_j}}}{{{k^2}}})\frac{{d({\omega ^2}{\varepsilon
_t}(\omega ,\vec k))}}{{d\omega }} \nonumber \\
& +& \frac{{{k_i}{k_j}}}{{{k^2}}}\frac{{d(\omega {\varepsilon
_l}(\omega ,\vec k)}}{{d\omega }}]{E_i}{E_j} \label{a}
\end{eqnarray}
 The field energy can be interpreted as a sum of
free harmonic oscillators provided  that longitudinal and transverse components of the field are multiplied by  respective  normalization factors ${N_l}^{-1} =
{\frac{k^2}{\omega }\frac{d}{d\omega }(\omega {\varepsilon _l})}
|_{\omega= \omega _{\vec k}}$, ${N_t}^{-1} = {{\frac{d}{{d\omega
}}({\omega ^2}{\varepsilon _t})}}|_{\omega=\omega _{\vec k}}$ \cite{harris}.
 Using the dispersions relations we may further simplify the factors to    $ {N_l}^2 =
\frac{\omega_p}{2k^2}(1-\frac{6k^2v_T^2}{\omega_p^2} )$, ${N_t}^2
= \frac{1+v_T^2}{2\omega } $. Field quantization is then implemented via  equal time commutation relations

$ \left[
A_i(\vec x,t),\prod_j(\vec y,t) \right] = \delta
_{ij}\delta ^{( 3)}\left( \vec x - \vec y
\right)
$, where $\prod_i(\vec r,t)=\int{d^3{r^\prime} dt^\prime}{\varepsilon_{ij}(t-t^\prime,\vec r-\vec r^\prime)\partial_t^{\prime} A_j(t^\prime,r^\prime)}$
The plane wave expansion for the photons given by
\begin{eqnarray}
A_i(x)=\int {\frac{{d^3}k}{(2\pi)^3}\sum\limits_{\lambda  = 1}^3
{{N^\lambda }(k)} (a_{k\lambda }{e^{ - ik.x}}{{\bf{\varepsilon} }_i^\lambda } + {{a^\dag }_{k\lambda
}}{e^{ik.x}}{{{{\varepsilon}^\ast }_i^\lambda}})}
\end{eqnarray}
where, $\lambda =$ 1,2 stand for transverse polarizations and $\lambda =$ 3 denotes the longitudinal polarization. With an explicit employment of $N_{l,t}$ leads to  the standard commutation relations
 $[a_{\vec{k} \lambda},
 a_{\vec{k}^{\prime}\lambda^{\prime}}^{\dagger}] = \delta(\vec{k} - \vec{k}^{\prime})\delta_{\lambda, \lambda^{\prime}}$.

Once the quantization rules are established,  it is straight forward to write Feynman rules  (which we shall not do here), and also the amplitudes for any process. The relevant process here is Compton scattering  $\gamma(\omega_i,\vec k_i,\alpha_i)+e^-(E_i, \vec p_i, s_i)
\rightarrow\gamma(\omega_f, \vec k_f, \alpha_f)+ e^-(E_f,\vec
P_f,s_f)$ where  $(\omega, \vec k, \alpha)$ and $(E, \vec p, s)$  refer to  energy, momentum and polarization for the photon and the electron respectively. Crucially, Compton scattering in the plasma medium  involves longitudinal polarization of the photon in addition to the transverse modes.
Accordingly, the scattering cross section is an incoherent  sum of four contributions coming from the longitudinal $(L)$ and transverse $(T)$ modes in initial and final states. The spin averaged/summed scattering cross section
for the process is given by
$d\sigma=\sum_{i,f} d\sigma_{if};~ i,f = L,T$.
 The target  electrons belong to the plasma at a definite temperature, and this calls for a further averaging of the cross section wrt to the initial electron distribution. All the results presented in this paper are reported after such an average. Thus the medium plays a dual role of (i) modifying the dispersion relations and the scattering amplitude through its permittivity and (ii)  further contributing to the average cross section. The role of the collective nature of the plasma is central to the analysis. It is pertinent to remark that the density and the temperature of the plasma have a highly nontrivial role to play.

The basic quantities of interest are  angular regions with momentum gain for the scattered electrons,  the angular distribution $\langle dN/d(\cos \theta_e)\rangle$   and the corresponding  energy spectrum $\langle dN/dE \rangle$; all of them are determined experimentally.  The momentum gain can be determined from dispersion relations, and the other two quantities by multiplying the corresponding cross sections by the initial photon flux. It is also of great interest to estimate yet another important quantity, the equivalent  electric field that causes the acceleration, since it is a measure of giant acceleration gradients which characterize all plasma based accelerators. We present the results below.

\begin{figure}
  \begin{center}
    \includegraphics[clip,width=\hsize]{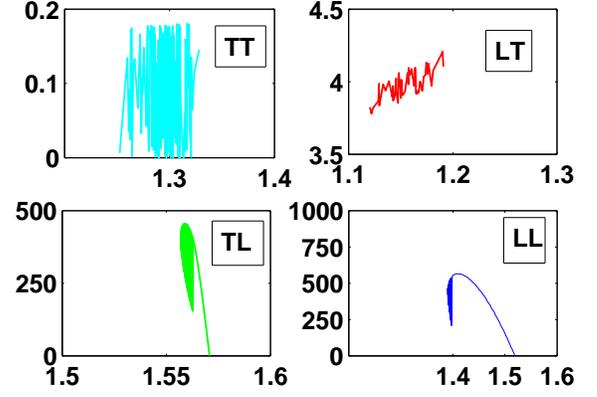}
    \caption{
    Angular regions of positive momentum transfer  to the electron (in $eV$). The scattering angle $\theta_e$ is defined wrt the beam direction.}
 \label{momentum_all1}
  \end{center}
\end{figure}

\noindent{\bf Results and Discussion}: The plasma parameters will be the same as chosen in the experiments.   We take the plasma density  $n_0 =6.5\times10^{18} cm^{-3}$ and  temperature $T_0= 50$ eV. The initial photon frequency $\omega_i=1.5$$\omega_p $, corresponding to a wavelength $\lambda = 9.31 \mu m$. The beam intensity is taken to be $I= 3.2 \times 10^{18} W cm^{-2}$.
 First of all, consider the momentum transferred to the electron in each case as the scattering angle is varied: Fig.\ref{momentum_all1} shows  regions
of positive momentum gain for each of the scattering processes. The angular width is very small, with $\Delta \theta \sim 3-10 ~~ mrad$ with the TL and LL cases having the least divergence. We further note
 that the dominant contribution to $\Delta P$ comes from $TL$ and $LL$ scatterings which can be $\sim 10^3$ larger than the other two cases.

\begin{figure}
  \begin{center}
    \includegraphics[clip,width=\hsize]{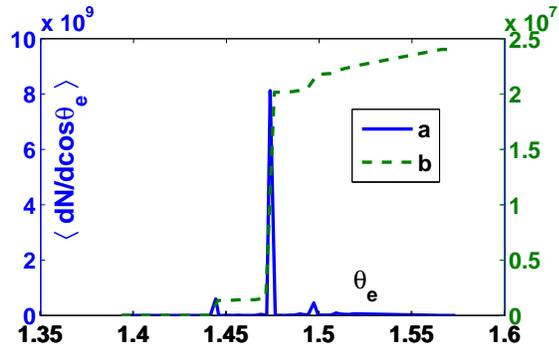}
\caption{ (a)Number density   and (b) the corresponding cumulative distribution of  scattered electrons as a function of $\theta_e$
}
 \label{dNdOmega_theta_e_10}
  \end{center}
\end{figure}
\begin{figure}
  \begin{center}
\includegraphics[clip,width=\hsize]{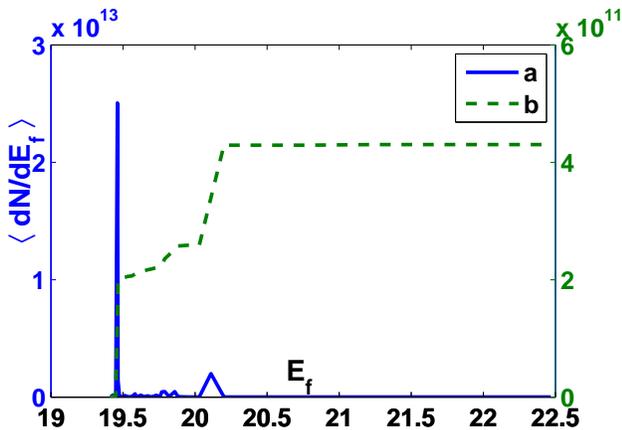}
\caption{ (a)Number density   and  (b) the cumulative distribution of the scattered electrons as a function of final electron energy $E_f$. Note the non Maxwellian nature.}
\label{dNdE_E_10}
\end{center}
\end{figure}

\begin{figure}
  \begin{center}
\includegraphics[clip,width=\hsize]{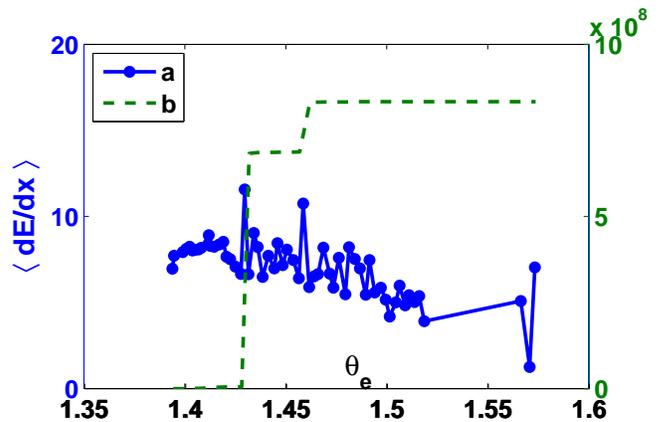}
\caption{ (a)Plot of $\log_{10} E_{eff}$, equivalent electric field (V/cm) as a function of  $\theta_e$. (b)
The corresponding cumulative distribution.}
\label{Efield_theta_e_10}
\end{center}
\end{figure}

\begin{figure}
  \begin{center}
    \includegraphics[clip,width=\hsize]{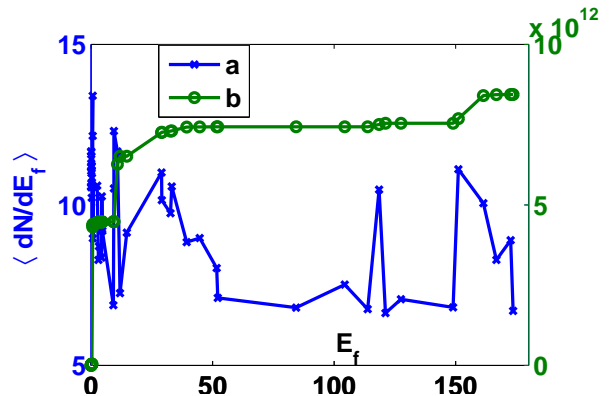}
\caption{ (a)Extrapolated curve for  $dN/dE_f$ after $10^7$ scatterings. (b)The corresponding cumulative distribution.}
 \label{dNdE_multi1_10}
  \end{center}
\end{figure}
\noindent {\bf Collimation and Energy Spectrum}:  We now  consider the number distribution $\langle dN/d(\cos \theta_e) \rangle$, in  Fig.\ref{dNdOmega_theta_e_10}, for the electrons scattered in the angular regions with the positive momentum gain  (See Fig. \ref {momentum_all1}); for clarity,   we show both the density as well as the cumulative distribution. Note again that the    contribution to $\langle dN/d(\cos \theta_e) \rangle$ is dominated  by  LL scattering  (Fig.\ref{momentum_all1}).
   Indeed,   the  cross sections $\Delta \sigma_{if}$ -- obtained by integration over the angular regions shown in Fig.\ref{momentum_all1} --  are respectively given by
$\Delta\sigma_{LL}=1.2\times10^9$, $\Delta \sigma_{TL}=2.1\times10^5$,
$\Delta\sigma_{LT}=1.3\times10^{-5}$, $\Delta\sigma_{TT}=0.0028$ in units of
$r_0^2=\alpha^2/m^2 = 0.08 \times 10^{-24} cm^2$, where $r_0$ is the classical electron radius.
These numbers may  be compared with the total cross section $\sigma_T \approx 10^{10}$ which is itself
 $~10^9$ times  larger than the Klien-Nishina cross section which is valid in free space.
  Fig.\ref{dNdOmega_theta_e_10} clearly demonstrates  collimation  which is restricted to a small angular width $\delta\theta \sim 2-3 \rm{mrad}$ centered at  $\theta_e= 1.474$. This is in excellent agreement with the experimental result of Geddes et al \cite{geddes} who report the same angular divergence, and is comparable to the findings of Faure et al \cite{faure} who report a divergence which is
 thrice the value found here. This establishes  the first goal that we set out to demonstrate.
 It is important to note that the angle at which  collimation takes place depends on  plasma parameters and the incoming beam energy. We report that we have  verified that  collimation is  stable against small perturbations in parameter values. Under large changes, e.g., in $n_0$, the beam quality is seen to deteriorate, as seen by experiments.
The energy distribution of the electrons in the angular region around collimating cone  is shown in Fig.\ref{dNdE_E_10}. The energy clearly peaks in the collimating cone, achieving a maximum density $dN/dE = 2.5   \times 10^{13}$ at $E=19.46 $ eV with a small width $\Delta E/E = 2.37\% $. Note further that the distribution is non Maxwellian, as reported in experiments\cite{faure,mangles,geddes}.\\
{\bf Equivalent Electric Field}: The equivalent electric field $E_{eff}$  in this formalism  is given by  $E_{eff}(\theta_e)=$
$\frac{1}{e}\langle\frac{d\mathcal{E}}{dx}\rangle (\theta_e)=\frac{n^{\gamma}_L}{e} \langle \Delta
E_L\frac{d\sigma}{d\Omega_e}\rangle$, where $n^{\gamma}_L $ is the mean photon density with longitudinal polarization and $E_L$ is the longitudinal component of the energy. Rather than  estimate $n^{\gamma}_l$ in terms of laser intensity and the medium properties, we take the value of $E_{eff}(\theta_e)$ as a phenomenological input from the experiments \cite{dt,faure}, who report a value  $\approx 10^9 V cm^{-1}$. This yields the value $n^{\gamma}_L \approx 10^{23} cm^{-3} $. At the collimating angle $\theta_e = 1.47$ rad, for the  $E_{eff} \approx  0.8 \times10^9 V cm^{-1}$ which compares excellently  with the estimate of \cite{dt} who find  $E_{eff} \sim 10^9 V cm^{-1}$. Employing this value, we plot $E_{eff}(\theta_e)$ in  Fig.\ref{Efield_theta_e_10} as a function of the scattering angle. The utility of this curve becomes clear below. \\
 \noindent {\bf Plasma Acceleration}: Finally,  we note that
  the energy of the scattered electron is peaked at ${\cal  E}_f \sim 20$ eV, which is  less than the thermal energy ($50$ eV). It might thus appear
   that the present effective field theory approach fails to capture the essential feature of the plasma acceleration, viz., very
   large energies. We now argue that this is not necessarily a draw back: only a single scattering has been considered here
   while in reality the final energy owes its value to multiple scatterings. To put this on a
 more quantitative  footing observe that the distance traversed by the electron due to a single scattering, $\ell=(n^{\gamma}_L\sigma)^{-1} \approx 0.3 $ \rm{nm} where $n^{\gamma}_L$ is taken from the above estimate. The number of scatterings in a distance of $\sim 3 $ mm is $O(10^7)$,
 which would contribute to a cumulative energy ${\cal E} \sim  100$ MeV.  which is not far away from the experimentally reported numbers. The result corresponding to this extrapolation is shown in Fig.\ref{dNdE_multi1_10} for electrons accelerated over a distance of a centimeter. The energy spectrum
  shows several peaks and is remarkably akin to the electron spectrum unraveled in \cite{faure} from deconvolution of their
  experimental data. Note however that our data gives two closely spaced prominent peaks, one at ${\cal E} \approx 118.5$ MeV
  which corresponds at a scattering angle $\theta_e = 1.45$ rad and the other at ${\cal E} \approx 150$  MeV at $\theta_e = 1.469$ rad.
  The second peak  is relevant since it occurs  in the collimating angular region.
  To conclude, we have seen that  the physics behind collimation,  quasi mono energetic spectrum and emergence of strong accelerating fields  in laser plasma accelerators  can  be understood in terms of an effective lagrangian. This may be employed  to study beat  wave acceleration and plasma wake field acceleration with equal ease. A careful evaluation of multi scattering cross sections and
  an explicit incorporation of the beam properties  would yield truly reliable quantitative estimates. Finally, the current approach allows a study higher order quantum corrections, similar to QED corrections to synchrotron radiation,  which are not easily accessible in  other methods.

 We are grateful to Vinod Krishan for bringing this problem to our notice. This work was supported by the Council of Scientific and Industrial
Research, New Delhi, India $ (09/092(0345)/2004-EMR-I)$


\begin{thebibliography}{99}
\bibitem{dt} J. M. Dawson and T. Tajima, Phys. Rev. Lett. 43, 267 (1979).
\bibitem{modena} A. Modena et al, Nature 337, 606-608 (1995).
\bibitem{umstadter} D. Umstadter et al, Science 273, 472 (1996).
\bibitem{moore} C. I. Moore et al, Phys. Rev. Lett. 79, 3909 (1997).
\bibitem{science} V. Malka et al, Nature 298, 1596 (2002).
\bibitem{faure} J. Faure et al, Nature 431, 541 (2004).
\bibitem{mangles} S. P. D. Mangles et al, Nature.431, 535 (2004).
\bibitem{geddes} C. G. R. Geddes et al, Nature 431,538 (2004).
\bibitem{malka} V. Malka et al, Phil. Trans. R. Soc. A.364, 601-610 (2006);
 T. Katsoules, Phys. Plasmas 13, 055503 (2006);
W. P. Leemans et al, Nat. Phys. 2, 418 (2006); C. Joshi, Phys. Plasmas 14, 055501 (2007).
\bibitem{muggli} P. Muggli et al, C. R. Physique, 10, 116-129, (2009);
 E. Esarey et al, Rev. Mod. Phys., 12, 1229 (2009);
 C. Joshi and V. Malka , New Journal of Physics, 12, 045003 (2010).
\bibitem{dawson} J. M. Dawson, Rev. Mod. Phys., 55, 403 (1983); W B Mori et al., Phys. Rev. Lett, 60, 1298 (1988);
 A. T. Lin, J. M. Dawson and H.Okuda., Phys.fluids. 17, 1995 (1974).
\bibitem{ghizzo} A. Ghizzo et al, J. Comput. Phys., 186, 47-69, (2003).
\bibitem{joshi2} C. Joshi et al., Phys. Rev. Lett.,  47, 1285, (1981).
\bibitem{landau}  Landau, Lifshitz, and Pitaevskii, 'Electrodynamics of continuous media', 1984,
                      2nd Edition Pergamon press. The normalization factors are derived in \cite{harris}.
 \bibitem{harris} Edward G. Harris,  'A Pedestrian Approach to Quantum Field Theory', 1972, John Wiley. 

\end{thebibliography}
\end{document}